# Investigation into U.S. Citizen and Non-Citizen Worker Health Insurance and Employment

By Annabelle Yao

The Lawrenceville School, 2500 Main Street, Lawrence Township, New Jersey, 08648,
ayhk2017@gmail.com

**Abstract**

Socioeconomic integration is a critical dimension of social equity, yet persistent disparities remain in access to health insurance, education, and employment across different demographic groups. While previous studies have examined isolated aspects of inequality, there is limited research that integrates both statistical analysis and advanced machine learning to uncover hidden structures within population data. This study leverages statistical analysis ($\chi^2$ test of independence and Two Proportion Z-Test) and machine learning clustering techniques—K-Modes and K-Prototypes—along with t-SNE visualization and CatBoost classification to analyze socioeconomic integration and inequality. Using statistical tests, we identified the proportion of the population with healthcare insurance, quality education, and employment. With this data, we concluded that there was an association between employment and citizenship status. Moreover, we were able to determine 5 distinct population groups using Machine Learning classification. The five clusters our analysis identifies reveal that while citizenship status shows no association with workforce participation, significant disparities exist in access to employer-sponsored health insurance. Each cluster represents a distinct demographic of the population, showing that there is a primary split along the lines of educational attainment which separates Clusters 0 and 4 from Clusters 1, 2, and 3. Furthermore, labor force status and nativity serve as secondary differentiators. Non-citizens are also disproportionately concentrated in precarious employment without benefits, highlighting systemic inequalities in healthcare access. By uncovering demographic clusters that face compounded disadvantages, this research contributes to a more nuanced understanding of socioeconomic stratification. These insights underscore the need for policies that expand health benefits equitably, regardless of citizenship status, and suggest that future studies should further explore targeted interventions to bridge gaps in both healthcare and employment protections.

**Keywords:** Statistics, Machine Learning, Immigration, Healthcare, Education,

1. **Introduction**

For many developed nations, pursuing economic stability and social integration for immigrant populations is a central challenge and priority. Access to stable employment and healthcare are two fundamental pillars of this integration process. However, significant gaps often exist between policy objectives and on-the-ground realities, potentially leading to systemic inequalities where certain groups face barriers to these essential services. Understanding the complex interplay between citizenship, employment, and healthcare access is a statistical exercise and a critical step toward crafting effective and equitable policy.

As of June 2024, over 19% of the US workforce—32 million out of a total of 169 million—are non-citizen workers and participate in the labor force at a higher rate than native-born workers, according to data from the Bureau of Labor Statistics (BLS)[1]. As such, the inequality of non-citizen worker treatment in areas of employment, wages, and healthcare are a large concern, especially in recent times.

This paper will leverage a χ2 test of independence to test for association between citizenship status and whether the workers worked for pay in the last week, and a two-proportion z-test at the 95% confidence level to check whether the proportion of citizen workers offered health insurance are greater than non-citizen workers who are offered health insurance. This study also investigates the nuanced relationship between citizenship status and socioeconomic outcomes by applying advanced unsupervised machine learning techniques to demographic and employment data.

Utilizing clustering algorithms—including K-Modes for categorical data and K-Prototypes for mixed data types—we segment the population into distinct profiles to uncover natural groupings that may be obscured by traditional analytical methods. The robustness of these clusters is then validated through t-SNE visualization for qualitative assessment and a CatBoost predictive model for quantitative evaluation. The primary objectives of this research are to determine whether citizenship status is a defining factor in workforce participation and to evaluate if disparities exist in access to employer-sponsored health benefits. By moving beyond correlation analysis to identify data-driven profiles, this research aims to reveal the underlying structure of socioeconomic integration and highlight areas where inequality may be systematically entrenched.

The dataset is the "2023 NATIONAL HEALTH INTERVIEW SURVEY (NHIS) (Version: 24 June 2024)"[2]. It comes from the government institutional records in the CDC (Centers for Disease Control and Prevention) National Center for Health Statistics, which is the principal source of information on the health of the civilian noninstitutionalized working population of the United States and is one of the major data collection programs of the NCHS initiated by the National Health Survey Act of 1956. The dataset used in this study consists of 29500+ individual data points analysed on 635 different categorical variables relating to the areas of healthcare, civic engagement, immigration status, wages, income, education, medicine, and more.

**1.1 Data Visualisation with TSNE**

tSNE has risen as a strong method of data visualisation. Originally created in 2008 by scientists Hinton and Van der Maaten to solve the shortcomings of existing techniques such as Principal Component Analysis (PCA), Sammon's Mapping, Isomap, and Locally Linear Embedding (LLE), tSNE alleviates the "crowding problem" plaguing the other techniques that expresses the impossibility to accurately represent both nearby and faraway distances in a low-dimensional space for complex data. tSNE is a nonlinear dimensionality reduction technique that projects high-dimensional data into a low-dimensional space[3]. The technique is a variation of Stochastic Neighbor Embedding (SNE) that is much easier to optimize, and produces significantly better visualizations by reducing the tendency to crowd points together in the center of the map[4]. SNE used a symmetric cost function (Kullback-Leibler divergence) that was difficult to optimize. It was susceptible to getting stuck in local minima and suffered from the crowding problem even more severely. It was also confusing to interpret because the probabilities were not symmetric.

However, tSNE solved the two major flaws of SNE. It uses a single, symmetric joint probability distribution in high-dimensional space, making the gradient of the cost function much simpler and faster to compute. Most importantly, instead of using a Gaussian distribution to calculate similarities in the low-dimensional map, t-SNE uses a heavy-tailed Student-t distribution (with one degree of freedom, essentially a Cauchy distribution)[5]. The heavy tails of the t-distribution allow points to be "pushed apart" more easily in the low-dimensional map. This dramatically alleviates the crowding problem, as the map now has much more space to organize moderately distant points.

The tSNE method functions significantly better than those produced by the other techniques such as Isomap and Locally Linear Embedding[3]. Over the years, more methods of tSNE data visualisation have arisen. One of which is using tree-based algorithms to accelerate tSNE. This method, through creating variants of the Barnes-Hut algorithm and of the dual-tree algorithm that approximate the gradient used for learning t-SNE embeddings in O(N log N), make it possible to learn embeddings of data sets with millions of objects[6].

**1.2 Supervised Learning**
Introduced in the mid-20th Century, supervised machine learning is the method in which the algorithm learns from a labeled dataset, meaning it learns from examples that include both input data and the corresponding correct output (or label). Through learning the relationships between the input and output, the algorithm is able to accurately predict or classify new, unseen data.

One of the most common types of supervised learning algorithms are decision trees. Decision tree represents a classifier expressed as a recursive partition of the instance space[7]. The decision tree consists of nodes that form a "root" tree, which means that it is a distributed tree with a basic node called root with no incoming edges. The core objective of a decision tree algorithm is to inductively learn a model from pre-labeled training data that can be used to make predictions on unlabeled instances. This learning process involves constructing a flow-chart-like structure that recursively partitions the feature space into purer subspaces, culminating in a predictive decision. Another method is linear regression[8]. Linear regression finds relationships and dependencies between variables through finding a single straight line that, on average, passes as closely as possible to a set of data points. It later uses that line to make predictions about new data.

Another prominent method of supervised learning is Catboost[9]. Catboost implements ordered boosting, a permutation-driven alternative to the classic algorithm, and has an innovative algorithm for processing categorical features. Together, these techniques help alleviate the problem caused by a prediction shift resulting from a special kind of target leakage present in all currently existing implementations of gradient boosting algorithms.

## 2. Methods
### 2.1 Research Design
To examine the relationship between citizenship status and employment-related outcomes, we conducted two statistical inference procedures and machine learning analysis using a dataset representing the population of interest, along with a randomly selected sample of 1,500 individuals.

### 2.2 Descriptive Statistics and Data Visualization
We first generated summary statistics for the full population to assess disparities in employment and health insurance access between citizens and non-citizens. To visualize these distributions, we constructed segmented bar charts comparing proportions across groups.

### 2.3 Chi-Square Test of Independence
We performed a $\chi^2$ test of association at the $\alpha = 0.05$ significance level to determine whether there was a statistically significant relationship between citizenship status and employment in the past week. The null hypothesis ($H_0$) stated that no association exists, while the alternative ($H_a$) posited an association. Before conducting the test, we verified all the conditions were satisfied.

### 2.4 Two-Proportion Z-Test
Next, we conducted a two-proportion z-test ($\alpha = 0.05$) to evaluate whether citizens were more likely than non-citizens to have been offered health insurance by their last employer. The hypotheses were:

$H_0: p_1 - p_2 = 0$ (no difference in proportions)

$H_a: p_1 - p_2 > 0$ (citizens have a higher proportion)

All conditions were checked.

### 2.5 Robustness Checks

Given the large sample size, we assessed whether the small p-value might reflect excessive statistical power rather than a meaningful effect. However, the population summary confirmed a substantial disparity, supporting the validity of our inference.

**2.6 K-modes**
K-modes is a clustering algorithm used in data science to group similar data points into clusters based on their categorical attributes. The algorithm extends the k-means clustering approach to handle categorical data by replacing the Euclidean distance with a dissimilarity measure for categorical attributes and using modes instead of means for cluster centroids[10].

Given 2 categorical variables X and Y with *m* features, the Hamming distance, which measures the number of positions at which the corresponding symbols are different (counting the minimum number of substitutions needed to transform one into the other) is:

$$d(X, Y) = \sum_{j=1}^{m} \delta(x_j, y_j)$$

Where:
$\delta(x_j, y_j) = 0 \quad if \quad x_j = y_j \ (same\ category)$
$\delta(x_j, y_j) = 1 \quad if \quad x_j \neq y_j \ (different\ category)$

The mode, the most frequent category, for each feature in a cluster is computed as:
$$Mode\ (C_k) = [mode(C_{k1}), mode(C_{k2}), mode(C_{k3})..., mode(C_{km})]$$
Where $C_k$ is the *k*-th Cluster. The algorithm minimizes the total cluster dissimilarity:

$$J = \sum_{k=1}^{K} \sum_{X \in C_k} d(X, Mode(C_k))$$

First, the algorithm is initialised by randomly selecting K initial modes from the data objects. Each object is assigned to the cluster with the nearest mode according to the minimum Hamming distance computed. The modes are then updated using the frequency-based method on newly formed clusters. Similarities between the data objects and updated modes are then recalculated. The steps are repeated until cluster assignments stabilize.

This paper uses k-modes to cluster the data and find groupings for the dataset.

**2.7 K-prototypes**
The k-prototypes algorithm generalizes k-means and k-modes to handle mixed data types (numerical + categorical)[11]. It combines the Euclidean distance for numerical features, the Hamming distance for categorical features and a weighting parameter $\gamma$ that balances the two distances. The algorithm groups the dataset into K clusters by minimizing the cost function:

$$E(U, Q) = \sum_{l=1}^{k} \sum_{i=1}^{n} u_{il} d(x_i, Q_l)$$

Where $Q_l$ is the the prototype of the cluster l; $u_{il}$ ($0 < u_{il} < 1$) is an element of the partition matrix $u_{n \times k}$; and $d(x_i, Q_l)$ is the dissimilarity measure which is given as:

$$d(x_i, Q_l) = \sum_{j=1}^{m} d(x_{ij}, q_{lj}),$$

$d(x_i, Q_l) = (x_{ij} - q_{lj})^2$ if the lth attribute is the numeric attribute,

$d(x_i, Q_l) = \mu_l \delta(x_{ij}, q_{lj})^2$ if the lth attribute is the categorical attribute.

Where $\delta(p, q) = 0$ for $p = q$ and $\delta(p, q) = 1$ for $p \neq q$; $\mu_l$ is a weight for categorical attributes in the cluster l. When $x_{ij}$ is a value of the numeric attribute, $q_{ij}$ is the mean of the jth numeric attribute in the cluster l; when $x_{ij}$ is the value of a categorical attribute, $q_{ij}$ is the mode of the jth categorical attribute in the cluster l.

The k-prototypes algorithm randomly chooses k data objects from the dataset X as the initial prototypes of clusters. For each data object in X, the algorithm assigns it to the cluster whose prototype is the nearest one to this data object in terms of either Hamming's distance or Euclidean distance. Following each assignment, the prototype of the cluster is updated. The similarity of data objects against the current prototypes after all data objects have been assigned to a cluster is recalculated. If a data object whose nearest prototype belongs to another cluster rather than the current one is discovered, reassign this data object to that cluster and update the prototypes of both clusters. After a full circle test of X, the algorithm ends if no data objects have changed clusters.

**2.8 t-SNE**
t-distributed stochastic neighbor embedding (t-SNE) is a nonlinear dimensionality reduction technique that projects high-dimensional data into a low-dimensional space (typically 2D or 3D) while preserving local structures and revealing underlying patterns[3]. It is particularly effective for cluster visualisation from algorithms such as k-modes or k-prototypes[6].

The high-dimensional space similarity is then increased if the probability of $x_j$ with $x_i$ as the center of the gaussian kernel is large. The denominator conducts normalization. For each pair of data points $x_i$ and $x_j$, the similarity $p_{j|i}$ in high dimensional space is computed as below

$$p_{j|i} = \frac{exp(-||x_i - x_j||^2 / 2\sigma_i^2)}{\sum_{k \neq i} exp(-||x_i - x_k||^2 / 2\sigma_i^2)}$$

In the low-dimensional space (with points $y_i$, $y_j$), similarities $q_{ij}$ are computed using a Student-t distribution.

$$q_{j|i} = \frac{(1+||y_i-y_j||^2)^{-1}}{\sum_{k \neq i}(1+||y_i-y_k||^2)^{-1}}$$

t-SNE minimizes the Kullback-Leibler (KL) divergence between the distributions P and Q:

Loss function

$$\arg\min D_{KL}(p||q) = \sum_i D_{KL}(p_i||q_i) = \sum_i \sum_j p_{j|i} \log \frac{p_{j|i}}{q_{j|i}}$$

Using the above as the cost function combined with stochastic gradient descent, we obtain the optimized solution:

$$\frac{\delta C}{\delta y_i} = 2\sum_j (p_{j|i} - q_{j|i} + p_{i|j} - q_{i|j})(y_i - y_j)$$

When converting high-dimensional data into conditional probabilities, it is necessary to set the standard deviation of the distribution centered on $x_i$. The adjustment method is to set the hyperparameter of perplexity. t-SNE will find the $P_i$ and standard deviation that matches the perplexity. The higher the perplexity setting, the greater the standard deviation will be. It is generally recommended that the value be set between 5 and 50.

$$Perp(P_i) = 2^{H(P_i)}$$

$$H(P_i) = -\sum_j p_{j|i} \log_2 p_{j|i}$$

**2.9 Catboost**
Catboost, also known as "Category Boosting", is an open-source gradient boosting library developed by Yandex[9]. Unlike other machine learning models that require categorical variables to be converted into numerical format through techniques such as one-hot encoding, CatBoost can work with these variables natively, allowing for a simplified data preparation process and enhanced model performance.

The choice of CatBoost was motivated by several key advantages over other gradient boosting frameworks (e.g., XGBoost, LightGBM)[12]:
- Native Handling of Categorical Features: CatBoost employs an efficient method of ordered boosting and a novel algorithm for processing categorical features, which eliminates the need for extensive pre-processing (e.g., one-hot encoding or label encoding) that can be computationally expensive and lead to information loss.
- Reduction of Target Leakage and Prediction Shift: Unlike standard gradient boosting algorithms that use the same dataset to calculate gradients and train the model, leading to biased pointwise gradient estimates, CatBoost's ordered boosting technique uses a permutation-driven approach to compute gradients, significantly reducing overfitting and prediction shift.
- Robustness and Performance: CatBoost has demonstrated state-of-the-art performance on various public benchmarks, particularly on datasets with numerous categorical features, and requires less hyperparameter tuning to achieve robust results.

CatBoost is built upon the gradient boosting decision trees (GBDT) framework[13]. The core principle involves building an ensemble of weak models (decision trees) in a sequential, greedy manner. Each subsequent tree is trained to correct the errors made by the previous ensemble. Given a training dataset with N samples and M features, where each sample is denoted as (x_i, y_i), as x_i is a vector of M features and y_i is the corresponding target variable, CatBoost aims to learn a function F(x) that predicts the target variable y. This can be expressed mathematically as such[14].

$$F(x) = F_0(x) + \sum_{m=1}^{M} \sum_{i=1}^{N} f_m(x_i)$$

where,
- $F(x)$ represents the overall prediction function that CatBoost aims to learn. It takes an input vector x and predicts the corresponding target variable y.
- $F_0(x)$ is the initial guess or the baseline prediction. It is often set as the mean of the target variable in the training dataset. This term captures the overall average behavior of the target variable.
- $\sum_{m=1}^{M}$ represents the summation over the ensemble of trees. $M$ denotes the total number of trees in the ensemble.
- $\sum_{i=1}^{N}$ represents the summation over the training samples. $N$ denotes the total number of training samples.
- $f_m(x_i)$ represents the prediction of the m-th tree for the i-th training sample. Each tree in the ensemble contributes to the overall prediction by making its own prediction for each training sample.

The equation states that the overall prediction $F(x)$ is obtained by summing up the initial guess $F_0(x)$ with the predictions of each tree $f_m(x_i)$ for each training sample. This summation is performed for all trees (m) and all training samples (i)[15].

## 3. Results

The summary statistics of the two categorical variables used in inference procedures are displayed in the tables below, using the dataset as our population, and a random sample of 1500 as our sample data. Table 1 shows the summary statistics for the variable EMPLASTWK_A (if the individual worked for pay last week) according to CITZNSTP_A (citizenship status) for the entire population.

|  | Worked for pay in the last week | | Did not work for pay in the last week | | |
|---|---|---|---|---|---|
|  | Frequency (people) | % | Frequency (people) | % | Total Frequency |
| Citizen | 14360 | 54.58% | 11950 | 45.42% | 26310 |
| Non-Citizen | 1256 | 65.93% | 649 | 34.07% | 1905 |

Table 1. Whether citizens and non-citizens worked for pay last week.

There is a significant disparity between citizens and non-citizens regarding whether or not they worked for pay in the last week. A slightly larger percentage of citizens worked for pay in the previous week (approximately 54.58%), compared to the 45.42% who didn't. Similarly, more non-citizens worked for pay last week than those who didn't. There is a 31.86% difference between the 65.93% of non-citizens who worked for pay the previous week and the 34.07% who didn't, with a greater percentage working for pay in the last week. This reflects that a large majority of non-citizens within the population worked for pay in the previous week, a trend echoed in citizens as well, with just over half working for pay in the last week.

Table 2 shows the summary statistics for EMPHEALINS_A (whether their last job offered health insurance) according to CITZNSTP_A (citizenship status).

|  | Health Insurance offered by their last job | | Their last job did not provide Health Insurance | | |
|---|---|---|---|---|---|
|  | Frequency (people) | % | Frequency (people) | % | Total Frequency |
| Citizen | 11644 | 71.65% | 4607 | 28.35% | 16251 |
| Non-Citizen | 760 | 53.45% | 662 | 46.55% | 1422 |

Table 2. Whether citizens and non-citizens were offered health insurance by their last job

There is a significant disparity between citizens and non-citizens regarding the availability of health insurance through their last job. Among citizens, 71.65% reported that their last job offered health insurance, while 28.35% indicated that their last jobs did not. There is a 43.3% difference between citizens between those who did and didn't have health insurance offered by their last job, reflecting a significant majority of citizens that benefit from such coverage. In contrast to this large majority among citizens, 53.45% of non-citizens had health insurance offered by their last job, and 46.55% of non-citizens did not have this opportunity. There is a 6.9% difference between the non-citizens who did and didn't have insurance offered, revealing a notable imbalance, where only a little over half of non-citizens had access to employer-provided health insurance. It also shows that non-citizens are less likely—by about 18 percentage points—to have access to health insurance through their employment compared to citizens. This will be shown in the later section using a two-proportion z-test.

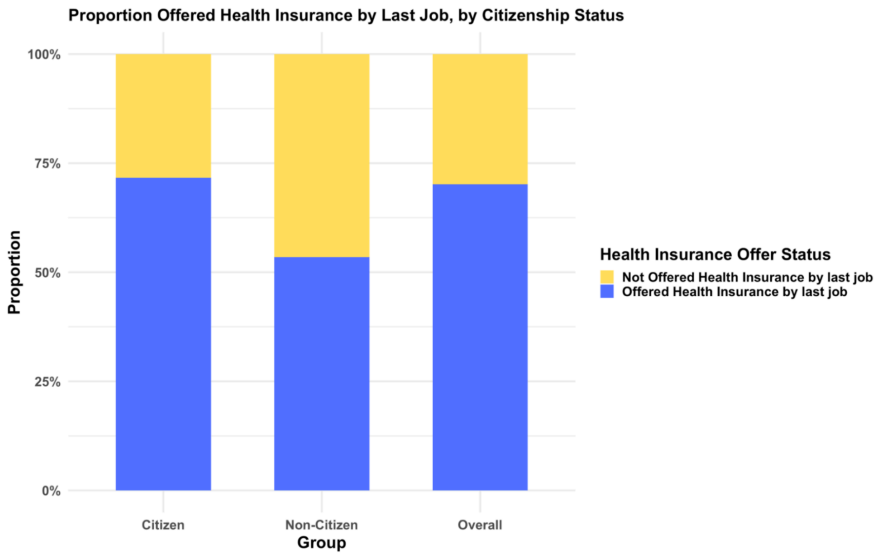

Figure 1. A segmented bar chart showing the proportion of offered health insurance by their last job according to citizenship status

Figures 1 and 2 show segmented bar charts for the population's EMPHEALINS_A and EMPLASTWK_A variables for easy distribution visualisation. Figure 1 shows the proportion of citizens, non-citizens, and the overall population regarding whether their last job offered health insurance. In the overall population, most people responded that they were offered Health Insurance at their last job, with around 70% saying they were offered, and over 25% saying they were not. The chart shows that non-citizens had the lowest proportion of people provided health insurance by their last job at just over 50%, while over 60% of citizens were provided health insurance by their last job.

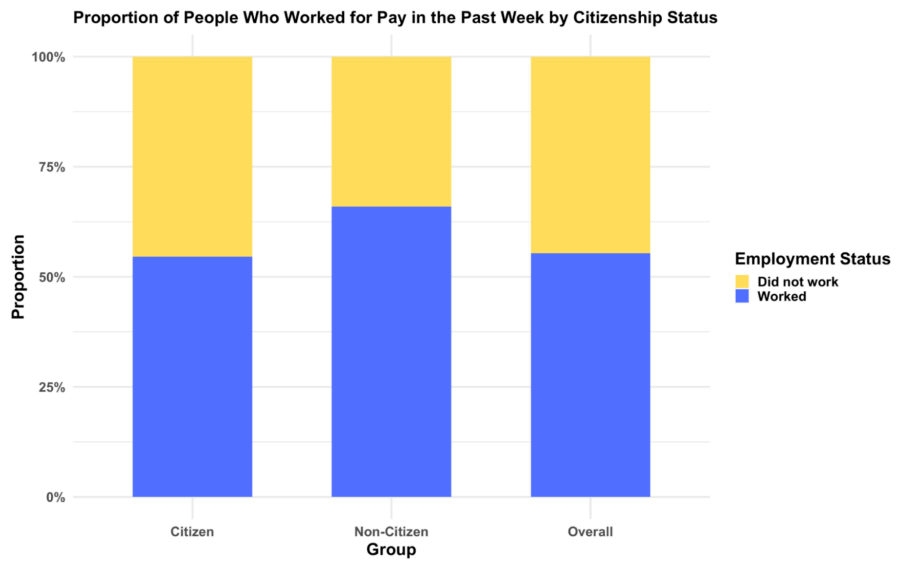

Figure 2. A segmented bar chart showing the proportion of people who worked for pay in the past week according to citizenship status

Figure 2 shows the distribution of responses to whether the people questioned worked for pay in the past week, revealing that the majority of the citizen, non-citizen, and overall population worked for pay in the past week. The chart also shows that the non-citizen group worked the most out of the other two, with over 63% working compared to just over 50% for citizens and the overall population.

### 3.1 χ2 Test of Association Between Working for Pay and Citizenship Status

Then, we conduct the χ2 test of association/independence at the α = 0.05 level on the randomly chosen 1500 samples. The null hypothesis is $H_0$: There is no association between whether individuals worked for pay last week and their citizenship status. The alternate hypothesis is $H_a$: There is an association between whether individuals worked for pay last week and their citizenship status.

The sample data of 1500 is shown in the table below.

|  | Worked for pay in the last week | Did not work for pay in the last week |
| --- | --- | --- |
| Citizen | 1217 | 164 |
| Non-Citizen | 109 | 10 |

Table 3. Contingency table for sample data of 1500 addressing citizens and non-citizens on whether or not they worked for pay in the last week

Before conducting the test, we check the conditions to ensure that the test can be done:
1. Random—satisfied; the 1500 samples are randomly chosen from the larger dataset
2. Expected Counts for each category ≥ 5—satisfied; 1221, 160, 105, 14 are all ≥ 5.

|  | Worked for pay (Expected) | Did not work for pay (Expected) |
| --- | --- | --- |
| Citizen | 1221 | 160 |
| Non-Citizen | 105 | 14 |

Table 4. Contingency table for expected sample data of 1500 addressing citizens and non-citizens, and whether or not they worked for pay in the last week

3. n ≤ (10%)N—satisfied; N ≈ 29500; (0.1)(29500) = 2950; 1500 ≤ 2950. The sample size is less than or equal to 10% of the population size, showing that the independence condition is met, which allows for the use of the inference procedure without introducing significant bias from sampling without replacement.
The χ2 test yields a p-value of 0.2564, a degree of freedom of 1, a χ2 value of 1.288.

Testing the hypothesis at the 5% significance level of α = 0.05. Since our p-value = 0.2564, significantly larger than α = 0.05, we fail to reject the null hypothesis; therefore, there is no convincing statistical evidence of an association between whether individuals worked for pay last week and their citizenship status.

### 3.2 Two-Proportion Z-Test for Health Insurance and Citizenship Status

Then, we use a two-proportion z-test to determine at the α = 0.05 whether a significantly greater proportion of citizens were offered health insurance from their previous job than non-citizens who were not. Our null hypothesis $H_0: p_1 - p_2 = 0$. Our alternate hypothesis $H_a: p_1 - p_2 > 0$. Where $p_1$ is the proportion of citizens who were provided health insurance from their previous job and $p_2$ is the proportion of non-citizens who were provided health insurance from their previous job. The summarized results from the randomly sampled 1500 individuals from the population are in the contingency table below.

|  | Health Insurance offered by their last job | Health Insurance not provided by their last job | total |
|---|---|---|---|
| Citizen | 1008 | 380 | 1388 |
| Non-Citizen | 63 | 49 | 112 |

Table 5. Contingency table for sample data of 1500 addressing citizens and non-citizens on whether their last job offered them health insurance.

Before conducting the test, we check the conditions for inference:
1. Random—satisfied; the 1500 samples were randomly chosen from the larger population.
2. np, n(1-p) ≥ 10 for $p_c$—satisfied.

$$p_c = \frac{x_1 + x_2}{n_1 + n_2} \Rightarrow p_c = \frac{1008 + 63}{1388 + 112} = 0.714$$

- $n_1 = 1388 \Rightarrow n_1 p_c \approx 991; n_1(1 - p_c) \approx 397; 991 \geq 10; 397 \geq 10$
- $n_2 = 112 \Rightarrow n_2 p_c \approx 80; n_2(1 - p_c) \approx 32; 80 \geq 10; 32 \geq 10$

3. n ≤ (10%)N—satisfied; N ≈ 29500; (0.1)(29500) = 2950; 1500 ≤ 2950. The sample size is less than or equal to 10% of the population size, showing that the independence condition is met, which allows for the use of the inference procedure without introducing significant bias from sampling without replacement.

We get a p-value of 0.00011 and a z-score of 3.6884. Since our p-value is 0.00011, which is significantly smaller than α = 0.05, we reject the null hypothesis, therefore there is convincing statistical evidence that the true proportion of all citizens who were offered health insurance from their previous job is higher than the true proportion of all non-citizens who were offered health insurance from their previous job.

Large datasets tend to produce small p-values because increased sample size will reduce variability. Thus, the standard error (S.E.), which has the large sample size n in the denominator, will decrease, making the sampling distribution of the sample proportion p-hat tighter around the true proportion value p. Since decreased S.E. is in the denominator of the z-statistic calculation, it will produce a larger Z-statistic, and therefore, a smaller p-value.

To check that there was no Type 1 error made in this inference procedure, this paper references the initial population summaries, where the proportion of citizens offered health insurance from their last job was 71.65%. In comparison, the proportion of the non-citizens provided was only 53.45%. Since the proportion of citizens and non-citizens offered health insurance from their last job is not equal, it can be confirmed that the null hypothesis (no difference in proportions) was correctly rejected, indicating the observed difference is unlikely due to random chance.

### 3.3 Identification of Population Segments via K-Modes and K-Prototypes Clustering

To further analyze the dataset, this paper has used K-modes and K-prototypes to split the dataset into clusters accordingly. Looking at the K vs. Clustering Cost graph below in Figure 3 that models the optimal number of clusters (k) in the clustering algorithm, it can be observed that the point where the rate of decrease sharply bends and becomes more gradual is around 4-5 clusters.

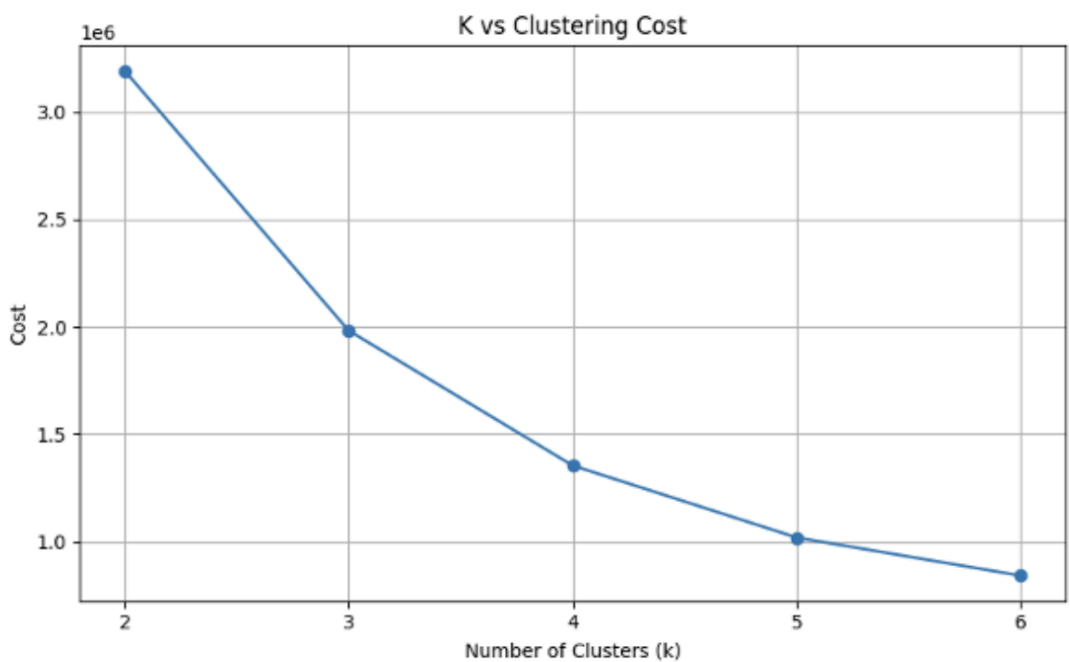

Figure 3. A K vs Clustering Cost chart showing the optimal number of clusters (4-5) for the k-prototypes clustering algorithm.

The distribution of each variable for each cluster can show the typical groups the dataset clusters into, thus revealing information about the connection between variables. The K-Prototypes clustering algorithm, applied to the categorical (Table 9) and quantitative socioeconomic variables (EMPWKHRS3_A: Hours worked per week, EMPDYSMSS3_A: Days missed work, past 12 months), identified five distinct and robust clusters within the immigrant population. Each cluster's centroids have been displayed in the Table 6 below.

| Cluster | EMPWKHRS3_A (Hrs Worked/ Week) | EMPDYSMSS3_A (Days Missed/Year) | EDUCP_A (Education) | NOTCOV_A (Insurance) | EMPWRKLSW1_A (Works for Pay) | CITZNSTP_A (Citizen) | NATUSBORN_A (Born in the US) | EMPLASTWK_A (Worked Last Week) | EMPHEALINS_A (Emp. Insurance) | EMPSICKLV_A (Sick Leave) |
|---|---|---|---|---|---|---|---|---|---|---|
| 0 | 37.9 | 109.1 | 1 (Less than HS) | 2 (Covered) | 1 (Yes) | 1 (Yes) | 1 (Yes) | 1 (Yes) | 1 (Yes) | 1 (Yes) |
| 1 | 39.5 | 35 | 2 (HS Grad+) | 2 (Covered) | 1 (Yes) | 1 (Yes) | 1 (Yes) | 1 (Yes) | 1 (Yes) | 1 (Yes) |
| 2 | 39.5 | 2.1 | 2 (HS Grad+) | 2 (Covered) | 1 (Yes) | 1 (Yes) | 2 (No) | 1 (Yes) | 1 (Yes) | 1 (Yes) |
| 3 | 17.4 | 1.7 | 2 (HS Grad+) | 2 (Covered) | 2 (No) | 2 (No) | 2 (No) | 2 (No) | 2 (No) | 2 (No) |
| 4 | 56.9* | 2 | 2 (HS Grad+) | 2 (Covered) | 2 (No) | 1 (Yes) | 1 (Yes) | 2 (No) | 1 (Yes) | 1 (Yes) |

Table 6. Table showing the centroids for each cluster.

*Note on Cluster 4 Hours: A value of ~57 hours is anomalous for a cluster that is not working (EMPLASTWK_A=2). This suggests that a small subset of outliers may influence the centroid calculation or that the feature has a highly skewed distribution. The categorical values are more reliable for interpreting this cluster's profile.*

With Table 6, one can infer the following characteristics for each cluster as displayed in Table 7 below.

| Cluster | Profile Name | Key Characteristics |
|---|---|---|
| 0 | The Hardworking Challengers | They work near full-time but miss a catastrophic number of days and have paid sick leave for them. Despite being less educated, they have insurance and benefits. |
| 1 | The Healthy, Integrated Native | The benchmark is full-time, healthy, educated, U.S.-born, with all employment benefits. |
| 2 | The Healthy, Integrated Immigrant | Identical to Cluster 1 in all outcomes except for being foreign-born. Represents successful integration. |
| 3 | The Precarious Non-Citizen | Works for very few hours. Lacks citizenship and all employment benefits (no health insurance, no sick leave) despite being educated. |
| 4 | The Healthy Retiree     | Not seeking work. Healthy, educated, insured (likely through retirement/government programs), and has benefits from a previous career. |

Table 7. The table shows the typical characteristics of each cluster according to the K-prototype centroid values.

Similarly, the K vs Clustering Cost graph below in Figure 4 for K-modes shows that 4-5 is the optimal number of clusters.

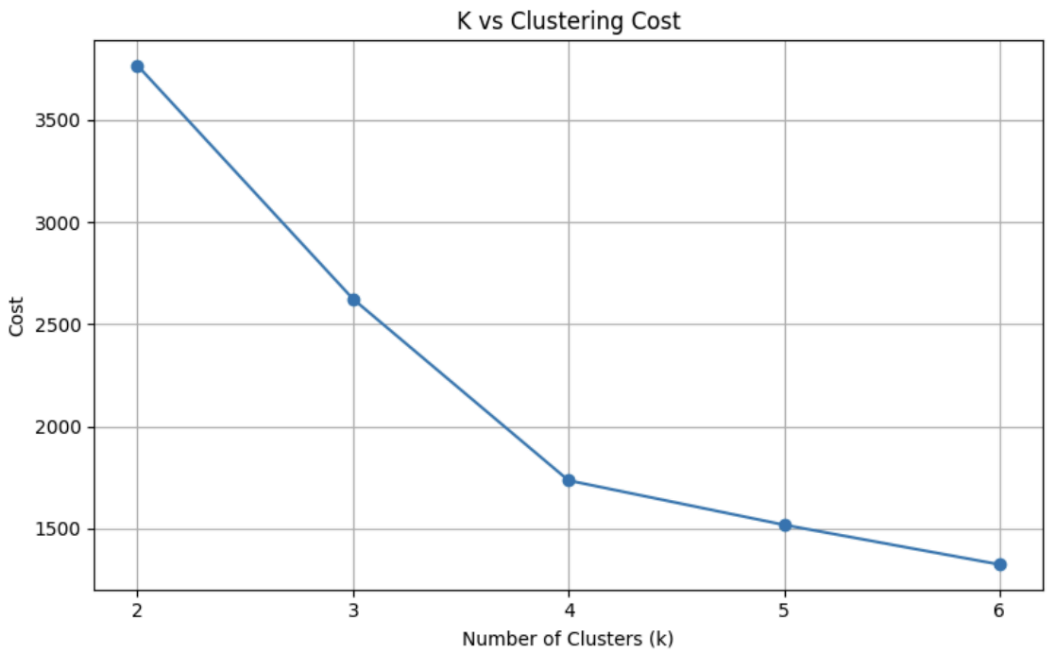

Figure 4. K vs Clustering Cost graph for K-Modes showing 4-5 as the ideal number of clusters

The created clusters each represent a group/type of people within the dataset. Specifically, the distribution of each variable according to cluster is shown in Table 8 below.

| Cluster | EDUCP_A | | NOTCOV_A | | EMPWRKLSW_A | | CITZNSTP_A | | NATUSBORN_A | | EMPLASTWK_A | |
|---|---|---|---|---|---|---|---|---|---|---|---|---|
| Y/N | 1 | 2 | 1 | 2 | 1 | 2 | 1 | 2 | 1 | 2 | 1 | 2 |
| 0 | 1 | 0 | 0.15 | 0.85 | 1 | 0 | 0.89 | 0.11 | 0.8 | 0.2 | 0.95 | 0.05 |
| 1 | 0 | 1 | 0.04 | 0.96 | 0.06 | 0.94 | 0.95 | 0.05 | 0.86 | 0.14 | 0 | 1 |

| | | | | | | | | | | | |
|---|---|---|---|---|---|---|---|---|---|---|---|
| 2 | 0 | 1 | 0.06 | 0.94 | 1 | 0 | 0.63 | 0.37 | 0 | 1 | 0.95 | 0.05 |
| 3 | 0 | 1 | 0.03 | 0.97 | 1 | 0 | 1 | 0 | 1 | 0 | 1 | 0 |
| 4 | 1 | 0 | 0.07 | 0.93 | 0 | 1 | 0.94 | 0.06 | 0.86 | 0.14 | 0 | 1 |

Table 8. Table displaying the distribution of each data variable according to cluster

With the data dictionary description for each variable shown in Table 9.

| Variable Name | EDUCP_A | NOTCOV_A | EMPWRKLSW_A | CITZNSTP_A | NATUSBORN_A | EMPLASTWK_A |
|---|---|---|---|---|---|---|
| Variable meaning | Educational level of the sample adults | Coverage status in Health United States | Worked last week | Citizenship Status | Born in the U.S. or a U.S. territory | Worked for pay the previous week |
| Meaning of 1 | Grade 1-11 | Not Covered | Yes | Yes, a citizen of the United States | Yes | Yes |
| Meaning of 2 | 12th Grade, No Diploma | Covered | No | No, not a citizen of the United States | No | No |

Table 9. Table displaying the data dictionary for the variables used.

According to Table 8, the distribution of each variable for each cluster can show the typical groups the dataset clusters into, thus revealing information about the connection between variables. The K-Modes clustering algorithm, applied to the categorical socioeconomic variables, identified five distinct and robust segments within the immigrant population. The composition and defining characteristics of these clusters are summarized in Table 10 below.

| Cluster | Size | Interpretation ("Typical Member") |
|---|---|---|
| 0 | 1206 | This person has **less than a high school diploma** (Grade 1-11 or 12th grade, no diploma). They **worked in the last week**, with 95% having received pay. They face significant economic hardship, with **the highest uninsured rate** (15%) out of all clusters. They are most likely a **U.S. citizen born in the U.S.** |
| 1 | 944 | This person has **at least a high school diploma**. 94% **did not work last week**, and the 6% that did work did not receive pay for their work (reasons can include voluntary work, not paid due to exploitation, etc.) They have a very high rate of **health insurance coverage** (96%). Consequently, it can be assumed that these people do not work for money and are likely voluntarily out of work (i.e., homemakers, students, retirees). They are almost certainly a **U.S. citizen born in the U.S**. |
| **2** | 326 | This person has **at least a high school diploma** and 94% have **health insurance**. They are **actively employed** (worked last week) and 95% worked for pay. The defining feature of this cluster is that 100% of its members were **born outside the United States**. A **majority (63%) have obtained U.S. citizenship**, while a significant minority (37%) have not. |
| **3** | 1326 | This person represents the prototype of full socioeconomic integration. They are **educated with at least a high school diploma**, 97% have **health insurance**, are U.S. citizens**, were born in the U.S.**, and are **actively employed** (worked last week for pay). This is the largest and most homogeneous cluster. |
| **4** | 1198 | This person has **less than a high school diploma**. They **did not work last week**. However, this group has a high rate of **health insurance coverage** (93%). This suggests reasons like retirement or disability, allowing them healthcare insurance likely through means like Medicare or Medicaid. They are **primarily U.S. citizens born in the U.S.** |

Table 10. A Table displaying the information for each cluster's typical member (majority taken) according to Table 8,9 data

This segmentation provides a clear, demographic-focused view of the population, revealing a primary split along the educational attainment (EDUCP_A), separating Clusters 0 and 4 from Clusters 1, 2, and 3. Furthermore, labor force status and nativity serve as secondary differentiators, identifying vulnerable groups (Cluster 0), successfully integrated immigrants (Cluster 2), and the prototypical native-born group (Cluster 3).

**3.4 Predictive Validation and Enhanced Feature Analysis with CatBoost**
To validate the robustness and predictability of these clusters, a CatBoost classifier was trained to predict cluster membership based on the original features, including both categorical (Table 9) and quantitative features. The variables added were as follows:
- EMPWKHRS3_A: Hours worked per week
- EMPDYSMSS3_A: Days missed work, past 12 months
- EMPSICKLV_A: Paid sick leave (1: Yes, 2: No)
- EMPWRKFT1_A: Number of adults in the sample adult's family who are working full-time
- EMPHEALINS_A: Health insurance offered

The model achieved a test accuracy of 92.3%, demonstrating that the clusters are well-separated and highly predictable based on the underlying feature set. This high accuracy provides strong validation that the K-Modes algorithm identified meaningful, coherent population segments.

The feature importance output from the CatBoost model offers a deeper, more nuanced understanding of the factors driving segmentation than was possible with the categorical-only

K-Modes analysis. The feature importance of each variable has been displayed in Table 11 below.

| Order of Importance | Feature Id | Importances |
|---|---|---|
| 1 | EDUCP_A | 23.983092 |
| 2 | EMPHEALINS_A | 23.290963 |
| 3 | EMPWKHRS3_A | 12.916838 |
| 4 | EMPDYSMSS3_A | 11.027137 |
| 5 | EMPSICKLV_A | 10.577578 |
| 6 | CITZNSTP_A | 9.457916 |
| 7 | NATUSBORN_A | 4.324617 |
| 8 | EMPLASTWK_A | 2.675093 |
| 9 | EMPWRKFT1_A | 1.746765 |

Table 11. The feature importance output for each variable after Catboost training.

The top predictive features were: EMPHEALINS_A (Access to employer-provided health insurance), EDUCP_A (Educational attainment), EMPWKHRS3_A (Usual hours worked per week), and EMPDYSMSS3_A (Days missed work). The CatBoost results both confirm and extend the insights from the K-Modes analysis. It confirms that the high importance of EDUCP_A (education) aligns perfectly with the K-Modes result, where education was the primary split between clusters.

The CatBoost model also identifies EMPHEALINS_A (employer health insurance) as the most important predictor. This critical numerical feature was unavailable to the K-Modes algorithm. Its top ranking reveals that access to employer-sponsored benefits is an even more powerful determinant of an immigrant's cluster membership than education alone.

This finding explains the economic reality behind the clusters. Clusters 2 and 3 (Integrated Immigrant and Native) have high predicted rates of employer-sponsored insurance, facilitating their high coverage. Cluster 4 (Safety-Net Dependent) lacks this access but achieves coverage through government programs. Cluster 0 (Precariously Employed) is likely employed in jobs that do not offer health insurance (EMPHEALINS_A = No) and may not qualify for safety-net programs, leading to their high uninsured rate.

**3.5 Visual Validation of Cluster Separation via t-SNE**

The high accuracy of the CatBoost classifier quantitatively confirmed that the clusters are predictable. A t-Distributed Stochastic Neighbor Embedding (t-SNE) projection was applied to the data to provide a qualitative, visual validation of cluster cohesion and separation.

t-SNE is a non-linear dimensionality reduction technique ideal for visualizing high-dimensional data in a two-dimensional space while preserving the local structure and relative distances between points. The resulting plot is shown in Figure 5.

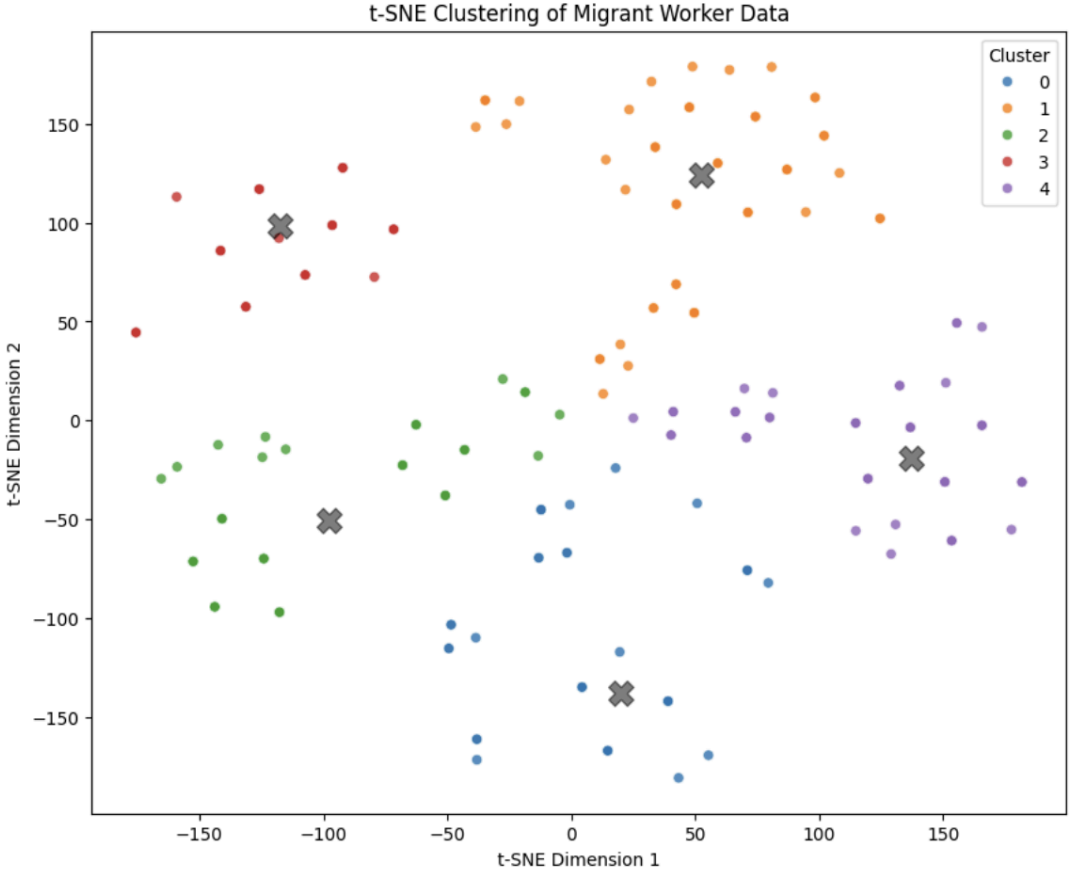

Figure 5. Graph of t-SNE clustering of Migrant Worker Data results. X marks the centroids of each cluster.

The t-SNE plot reveals several key findings that corroborate the quantitative results:
  (1) Clear Cluster Separation: The five clusters identified by the K-Modes algorithm form distinct, well-separated groupings in the two-dimensional embedding. This visual separation provides strong evidence that the clusters represent genuine, natural subgroups within the data and are not an artifact of the algorithm.
  (2) High Intra-Cluster Cohesion: Each cluster's data points are tightly grouped. This indicates that individuals assigned to the same cluster are highly similar to each other across the original high-dimensional feature space, confirming the internal homogeneity of each segment.
(3) Absence of Significant Overlap: The boundaries between clusters are notably clear, with minimal overlapping points from different groups. This visual finding aligns perfectly with the high (92.3%) predictive accuracy of the CatBoost model, as well-defined clusters are inherently easier for a classifier to learn and predict.
(4) Identification of Outliers: The plot shows a few isolated points far from any cluster centroid. This suggests that the chosen clustering solution effectively captures most of the population's structure, with few outliers that don't fit into a significant profile.

The t-SNE visualization serves as the final layer of validation, creating a powerful convergence of evidence: the K-Modes and K-prototypes algorithms found five demographic profiles, the CatBoost model proved these profiles are highly predictable and revealed that employment-based features are key differentiators. The t-SNE plot visually demonstrates that these five profiles are genuinely distinct and well-formed entities within the data's inherent structure.

## 4. Discussion

This comprehensive analysis combined a statistical analysis ($\chi^2$ test of independence and two-proportion z-test) with a multi-faceted machine learning approach—including K-Modes, K-Prototypes, t-SNE visualization, and CatBoost classification—to segment a population based on key socioeconomic and employment variables. The convergence of findings across these methods provides a robust and nuanced understanding of the underlying structure within the data.

While the findings reveal very meaningful patterns, they are also limited by the scope of the dataset and the associative nature of the analysis, suggesting the need for future research with broader data and causal frameworks.

Future work will focus on applying explainable AI (XAI) techniques, such as SHAP analysis, to interpret the CatBoost model's decisions and pinpoint the exact marginal impact of each feature on cluster assignment. Additionally, longitudinal tracking of these clusters could assess mobility between groups and evaluate the effects of policy changes on reducing observed inequalities.

In this research, the t-SNE visualization confirmed that the identified clusters are genuinely distinct, well-separated groups within the high-dimensional data space. The exceptional performance of the CatBoost classifier (92.3% accuracy) in predicting cluster membership further validates that these segments are not arbitrary but are defined by strong, predictable patterns in the data. The K-Modes analysis provided an initial demographic segmentation, which was then significantly enriched by the K-Prototypes model. The latter revealed that numerical features like hours worked and days missed due to illness are profound differentiators, uncovering a segment of individuals (Cluster 0) grappling with severe health crises despite being employed.

Most significantly, the K-Prototypes centroids precisely quantified the disparities inferred from the models. It can be conclusively stated that there was no association between citizenship status and labor force participation; individuals across citizenship statuses were equally present in the workforce. However, a stark contrast emerged in the quality of employment. A significantly higher proportion of citizens than non-citizens were offered health insurance and other employer benefits. This is unequivocally demonstrated by Cluster 3, a group defined by its non-citizen status, part-time precarious work, and—most critically—a complete lack of employer-provided health insurance and paid sick leave.

Together, these results reveal that systemic inequality persists not in the opportunity to work, but in the quality and security of that work. These findings illuminate the path to progress: society must move beyond ensuring mere employment and walk decisively toward building a truly just system where the foundational benefits of healthcare and economic stability are accessible to every worker, regardless of their origin or citizenship status.

## 5. Acknowledgements

Thank you to my school statistics teacher Mr Wilcox. Through applying the concepts learned in his class, I was able to produce this statistical analysis of Immigration data.

Thank you to my parents and my brother who have been incredibly supportive throughout this research process.

# References

[1] USAFacts. (2023, November 15). How many immigrants are in the American workforce? Retrieved May 16, 2025, from USAFacts website: https://usafacts.org/articles/how-many-immigrants-are-in-the-american-workforce/

[2] 2023 NATIONAL HEALTH INTERVIEW SURVEY (NHIS) (Version: 24 June 2024). (n.d.). *National Center for Health Statistics National Health Interview Survey*. Retrieved May 20, 2025, from https://www.cdc.gov/nchs/nhis/documentation/2023-nhis.html

[3] Van der Maaten, L., & Hinton, G. (2008). Visualizing Data using t-SNE Laurens van der Maaten. *Journal of Machine Learning Research*, 9, 2579–2605. Retrieved from https://www.jmlr.org/papers/volume9/vandermaaten08a/vandermaaten08a.pdf

[4] Hinton, G. E., & Roweis, S. (2025). Stochastic Neighbor Embedding. Advances in Neural Information Processing Systems, 15. Retrieved from https://papers.nips.cc/paper_files/paper/2002/hash/6150ccc6069bea6b5716254057a194ef-Abstract.html

[5] Arora, S., Hu, W., & Kothari, P. K. (2018). An Analysis of the t-SNE Algorithm for Data Visualization. *PMLR*, 1455–1462. Retrieved from https://proceedings.mlr.press/v75/arora18a.html

[6] Van der Maaten, L. (2014). Accelerating t-SNE using Tree-Based Algorithms Laurens van der Maaten. *Journal of Machine Learning Research*, 9, 2579–2605. Retrieved from https://www.jmlr.org/papers/volume9/vandermaaten08a/vandermaaten08a.pdf

[7] Lior Rokach, Oded Z. Maimon. (2008) *Data Mining with Decision Trees: Theory and Applications.* In World Scientific.

[8] C.-J. Lin, R. C. Weng, S. S. Keerthi. (2008) *Trust region Newton method for large-scale logistic regression.* In Journal of Machine Learning Research, vol. 9.

[9] Prokhorenkova, L., Gusev, G., Vorobev, A., Dorogush, A., & Gulin, A. (n.d.). CatBoost: unbiased boosting with categorical features. Retrieved from https://proceedings.neurips.cc/paper_files/paper/2018/file/14491b756b3a51daac41c24863285549-Paper.pdf

[10] Huang, Z. (1998). Extensions to the k-Means Algorithm for Clustering Large Data Sets with Categorical Values. Data Mining and Knowledge Discovery, 2, 283–304. Retrieved from https://cse.hkust.edu.hk/~qyang/537/Papers/huang98extensions.pdf

[11] Ji, J., Bai, T., Zhou, C., Ma, C., & Wang, Z. (2013). An improved k-prototypes clustering algorithm for mixed numeric and categorical data. *Neurocomputing, 120*, 590–596. https://doi.org/10.1016/j.neucom.2013.04.011

[12] Jeremiah, O. (2024, September 6). CatBoost in Machine Learning: A Detailed Guide. Retrieved August 22, 2025, from Datacamp.com website: https://www.datacamp.com/tutorial/catboost

[13] GeeksforGeeks. (2021, January 20). CatBoost in Machine Learning. Retrieved August 22, 2025, from GeeksforGeeks website: https://www.geeksforgeeks.org/machine-learning/catboost-ml/

[14] GeeksforGeeks. (2023, November 9). How CatBoost algorithm works. Retrieved August 22, 2025, from GeeksforGeeks website: https://www.geeksforgeeks.org/machine-learning/catboost-algorithms/

[15] Kolli, A. (2024, February 13). Understanding CatBoost: The Gradient Boosting Algorithm for Categorical Data. Retrieved August 22, 2025, from Medium website: https://aravindkolli.medium.com/understanding-catboost-the-gradient-boosting-algorithm-for-categorical-data-73ddb200895d